\documentclass[aps,twocolumn,showpacs,showkeys,amsmath,amssymb]{revtex4-1}
\usepackage{graphicx}

\begin{document}

\newcommand{\pderiv}[2]{\frac{\partial #1}{\partial #2}}
\newcommand{\deriv}[2]{\frac{d #1}{d #2}}

 \title{Role of conviction in nonequilibrium models of opinion formation}

\author{Nuno Crokidakis$^{1}$}
\thanks{E-mail: nuno.crokidakis@fis.puc-rio.br}

\author{Celia Anteneodo$^{1,2}$}
\thanks{E-mail: celia@fis.puc-rio.br}

\affiliation{
$^{1}$Departamento de F\'{\i}sica, PUC-Rio, Rio de Janeiro, Brazil \\
$^{2}$National Institute of Science and Technology for Complex Systems, Rio de Janeiro, Brazil}

\date{\today}

\begin{abstract}
\noindent
We analyze the critical behavior of a class of discrete opinion models in 
the presence of disorder. 
Within this class, each agent opinion takes a discrete value ($\pm 1$ or 0)  and 
its time evolution is ruled by two terms, one representing agent-agent interactions and 
the other the degree of conviction or persuasion (a self-interaction). 
The mean-field limit, where each agent can interact evenly with any other, is considered. 
Disorder is introduced in the strength of both interactions, with either 
quenched or annealed random variables. 
With probability $p$ (1-$p$), a pairwise interaction reflects a negative (positive) coupling, while 
the degree of conviction also follows a binary probability distribution (two different discrete 
probability distributions are considered). 
Numerical simulations show that a non-equilibrium continuous phase transition, 
from a disordered state to a state with a prevailing opinion, 
occurs at a critical point $p_{c}$ that 
depends on the distribution of the convictions,  the transition being spoiled in some cases. 
We also show how the critical line, for each model, is affected by the   
update scheme (either parallel or sequential) as well as by the 
kind of disorder (either quenched or annealed).

\end{abstract}




\maketitle

\section{Introduction}

In the last decades, diverse questions of social dynamics have been studied through 
statistical physics techniques. 
In fact, simple models allow to simulate and understand real problems such as 
elections, spread of information, vehicle traffic or pedestrian evacuation, 
amongst many others \cite{loreto_rmp}. 
As a feedback, these issues are attractive to physicists because of the occurrence of 
order-disorder transitions, scaling and universality, among other typical features of 
physical systems.

Concerning the particular subject of opinion dynamics, several models have been proposed in order to 
study the emergence of consensus    (for a recent review, see \cite{loreto_rmp}).
As concrete examples, let us mention opinion models based on outflow 
dynamics \cite{sznajd,sznajd_meu,sznajd_meu_pmco}, 
majority rules \cite{redner,galam_maj,galam,galam_epjb} 
and bounded confidence \cite{deffuant}, as well as kinetic 
exchange \cite{lccc,biswas,p_sen,biswas11}. 
Recently, the effects of negative interactions \cite{biswas,wang} 
and network dynamics \cite{newman,kozma,gross_review} in opinion formation have 
also been considered.

In this work we introduce  heterogeneity in  
the degree of persuasion or conviction of the agents. 
It is mimicked by a parameter which gauges the tendency of  an agent to hold its opinion 
or (if negative) change mind spontaneously.    
Then, we study the impact of persuasion 
in the critical behavior of a non-equilibrium model of opinion formation with a 
finite fraction of random negative agent-agent interactions.
We study two classes of disorder (either quenched or annealed) both for 
the strength of convictions and for agent-agent couplings. 
We also consider two different kinds of update, either sequential or parallel. 
Numerical Monte Carlo simulations  show that a continuous order-disorder 
phase transition, where order is characterized by a dominating opinion, 
can occur in all the variants of the model considered. 
However, the critical line is strongly affected by the distribution of convictions. 
Moreover, it is also affected both by the update scheme and 
by  the nature of the random variables, 
as occurs in other models 
\cite{stauffer,sabatelli,adriano_updates,bolle,schonfisch,caron}.

This work is organized as follows. In Section II we present the opinion model and 
define its microscopic rules. Numerical results are discussed 
in Section III in connection with the analytical considerations presented in the 
Appendix. Section IV contains the conclusions and final remarks.


\section{The model}

We consider an opinion model  based on kinetic exchange \cite{lccc,biswas,p_sen,biswas11}. 
At a given time step $t$, each agent $i$ has a discrete opinion $o_{i}(t)=-1, 0$ or $1$,   
that evolves according to
\begin{eqnarray}  
o_{i}(t+1) & = & C_{i}\,o_{i}(t) + \mu_{ij}\,o_{j}(t) ~ , \nonumber \\ 
o_{j}(t+1) & = & C_{j}\,o_{j}(t) + \mu_{ji}\,o_{i}(t)  ~, \label{ev_eq}
\end{eqnarray} 
where $C_{i}$ is the conviction of agent $i$ and $\mu_{ij}$ is the strength of 
the influence it suffers from  a randomly chosen agent $j$  in a fully-connected graph. 
If the value of the opinion exceeds (falls below) 
the  value 1 ($-1$), then it adopts the extreme value 1 ($-1$). 
Pairwise interaction strengths are random variables distributed 
according to the binary probability density function (PDF)
\begin{equation} \label{mudist}
F(\mu_{ij}) = p\,\delta(\mu_{ij}+1) + (1-p)\,\delta(\mu_{ij}-1) ~.
\end{equation}
In other words, the agents can exchange opinions with positive ($+1$) or negative ($-1$) influences, 
and $p$ quantifies the mean fraction of negative ones \cite{biswas}. 
In  magnetic systems, analogous positive (negative) interactions would 
correspond to ferro (anti-ferro) couplings. 
Notice certain similarities with what is known as the (mean-field) Blume-Capel  model \cite{blume_capel}:    
each opinion has three different states (spin-1 Ising);  
agents interact through ferromagnetic/anti-ferromagnetic couplings; 
in the Hamiltonian defining the model, there are quadratic terms 
representing the interaction of the spins with the crystal field 
and that can be related to the agents self-interaction; 
finally, the Blume-Capel model may include the interaction with an external field, 
that, although neglected here, may be opportune in opinion models as well, 
representing for instance propaganda or other external conditioning feature \cite{nuno_physicaA}.
Since in the present model, couplings are random:  
positive/ferromagnetic (or negative/anti-ferromagnetic) 
with probability $1-p$ (or $p$), it remits to the random-bond version of the Blume-Capel model, 
with random local competing fields.  
Moreover, here there is absence of thermal fluctuations, corresponding to the zero temperature 
limit of thermal spin models. Zero temperature random Ising-like models, for instance containing   
either local or global random fields, have already been considered 
to model group decision making \cite{decision,collective}. 
Notice however that differently from those magnetic models, the interactions occur by pairs 
and there is not an energy-like function to optimize. As a consequence of the different 
dynamical rules, the critical behavior is not related to 
that of usual equilibrium models, as we will see in the results presented in the next Section.  
For instance, no frozen or spin-glass phase is observed.  
The phenomenological differences were explained before as being 
due to the lack of frustration, despite the competitive random interactions, as soon as interactions 
do not occur simultaneously \cite{biswas}.

The influence of an individual over another one needs not be reciprocal 
(i.e., not necessarily $\mu_{ij}=\mu_{ji}$), however, 
whether interactions are symmetric or not, does not affect the results. 
If $C_{i}=1$ for all $i$ (i.e., $q=1$), one recovers the model of  Ref. \cite{biswas}, 
for which there is  
an order-disorder transition at a critical value $p_c=1/4$. 
As discussed in Ref. \cite{biswas}, the effect of negative interactions is similar 
to that produced by Galam's contrarians in opinion models \cite{galam_cont}. We will discuss this relation in more details in the following.

However, more realistically, the degree of conviction needs not be unitary nor homogeneous \cite{biswas11}. 
Then we considered two discrete alternatives for the PDFs of the convictions $C_{i}$, 
namely, 
\begin{eqnarray}  
G_1(C_{i})  &=&   q\,\delta(C_{i}-1) + (1-q)\,\delta(C_{i}-0)\,,   \label{Cdist1} \\ 
G_2(C_{i})  &=&    q\,\delta(C_{i}-1) + (1-q)\,\delta(C_{i}+1) \,.   \label{Cdist2}
\end{eqnarray}
They model the cases where a mean fraction $1-q$ of the individuals have either no convictions or 
completely change mind, respectively. 
In comparison to magnetic models, $G_1$ and $G_2$ are related  
to random diluted field and random antiferromagnetic impurities, respectively \cite{spin1BC}.

In both cases, the model of Ref. \cite{biswas} is recovered for $q=1$. 
Notice that, differently from the Sznajd dynamics \cite{sznajd}, where each agent 
interacts with a group 
of individuals at a time, in the present exchange model, interactions are pairwise.

We will show how the heterogeneity of convictions favors disorder 
or even provokes the destruction of the order-disorder phase transition.
Moreover, we will analyze two distinct kinds of the random variables $C_{i}$ and $\mu_{ij}$: 
they can be either quenched or annealed. The former are drawn from the PDFs given by 
 Eqs. (\ref{mudist}) and (\ref{Cdist1}) [or (\ref{mudist}) and (\ref{Cdist2})] 
at the beginning of each simulation and remain fixed during the evolution of the system, 
whereas the later are renewed at each  Monte Carlo step (MCs), where one MCs 
corresponds to $N$ iterations of Eq.~(\ref{ev_eq}), being $N$ the population size. 

In addition, we have studied two kinds of upgrades: synchronous (parallel) and 
asynchronous (random sequential). In the former case, we randomly choose 
$N$ pairs of agents that interact by means of Eq. (\ref{ev_eq}). 
Only after the $N$ interactions took place, the states of the $N$ agents are 
simultaneously renewed, increasing time by one  MCs. 
In the  asynchronous case, also  $N$ pairs of agents that interact by means 
of Eq. (\ref{ev_eq}) are randomly chosen at each MCs, 
but the opinions are assigned a new value at each interaction. 
A more realistic  dynamics  probably proceeds in between both schemes.

All simulations start with a random initial distribution 
of opinions, and all interacting pairs of agents are randomly chosen among the $N$ individuals 
in the population (which corresponds to a mean-field approach).


\section{Results}

We analyze the critical behavior of the system, in analogy to Ising spin systems, 
by computing the order parameter  
\begin{equation} \label{order}
O = \left\langle \frac{1}{N}\left|\sum_{i=1}^{N} o_{i}\right|\right\rangle ~, 
\end{equation}
where $\langle\, ...\, \rangle$ denotes disorder or configurational average. 
Notice that $O$ plays the role of  the ``magnetization per spin'' in  magnetic systems. 
In addition, we also consider the fluctuations $\chi$ of the order parameter (or ``susceptibility'') 
\begin{equation} \label{chi}
\chi =  N\,(\langle O^{2}\rangle - \langle O \rangle^{2})   
\end{equation}
and the Binder cumulant $U$, defined as  
\begin{equation} \label{binder}
U   =   1 - \frac{\langle O^{4}\rangle}{3\,\langle O^{2}\rangle^{2}} \,.
\end{equation}
 
In the following subsections, we will analyze separately 
the distributions given by Eqs. (\ref{Cdist1}) and (\ref{Cdist2}).

\subsection{Model with distribution $G_1$}

\begin{figure}[b]
\begin{center}
\vspace{0.4cm}
\includegraphics[width=0.35\textwidth,angle=0]{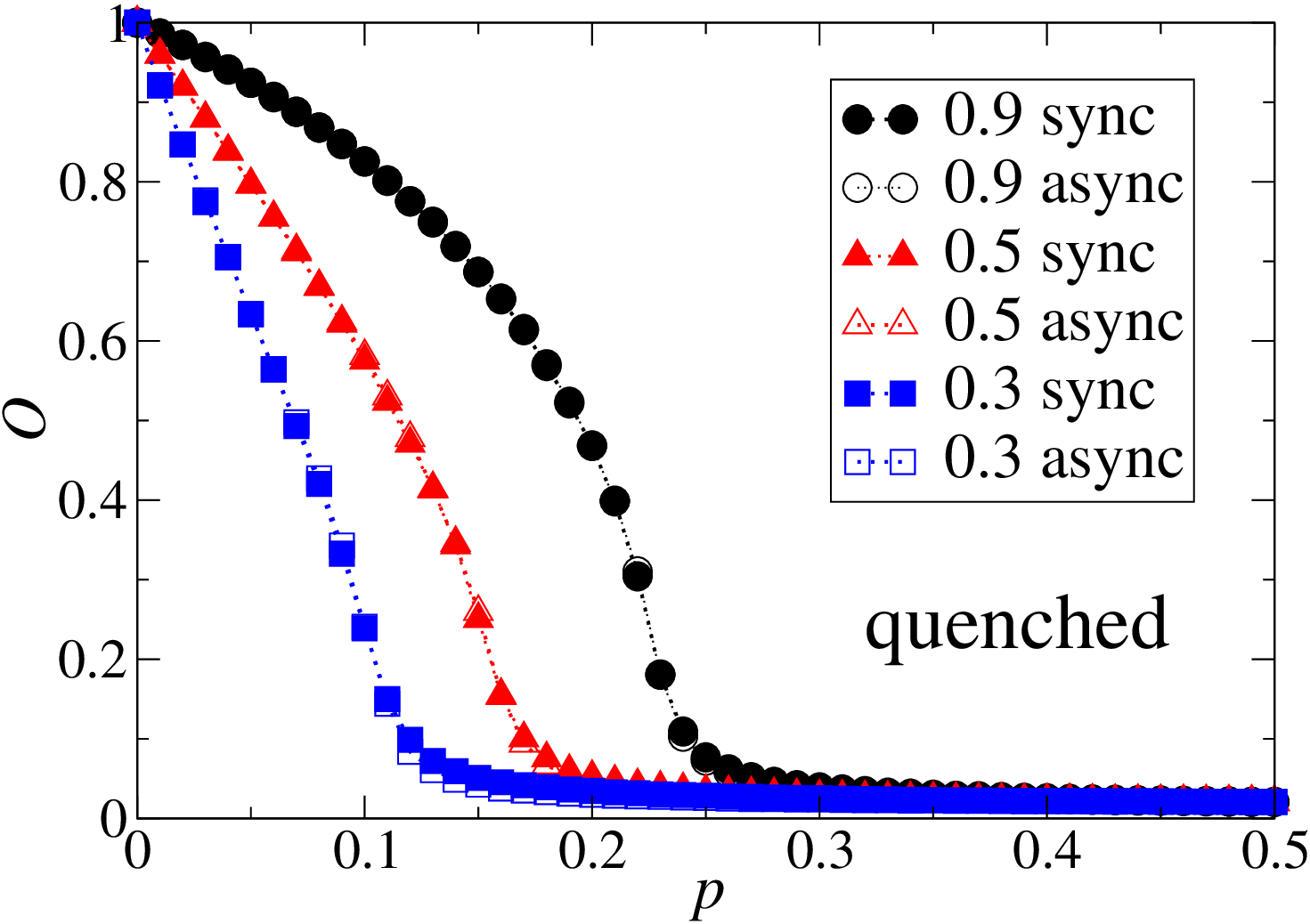} \\
\vspace{0.45cm}
\includegraphics[width=0.35\textwidth,angle=0]{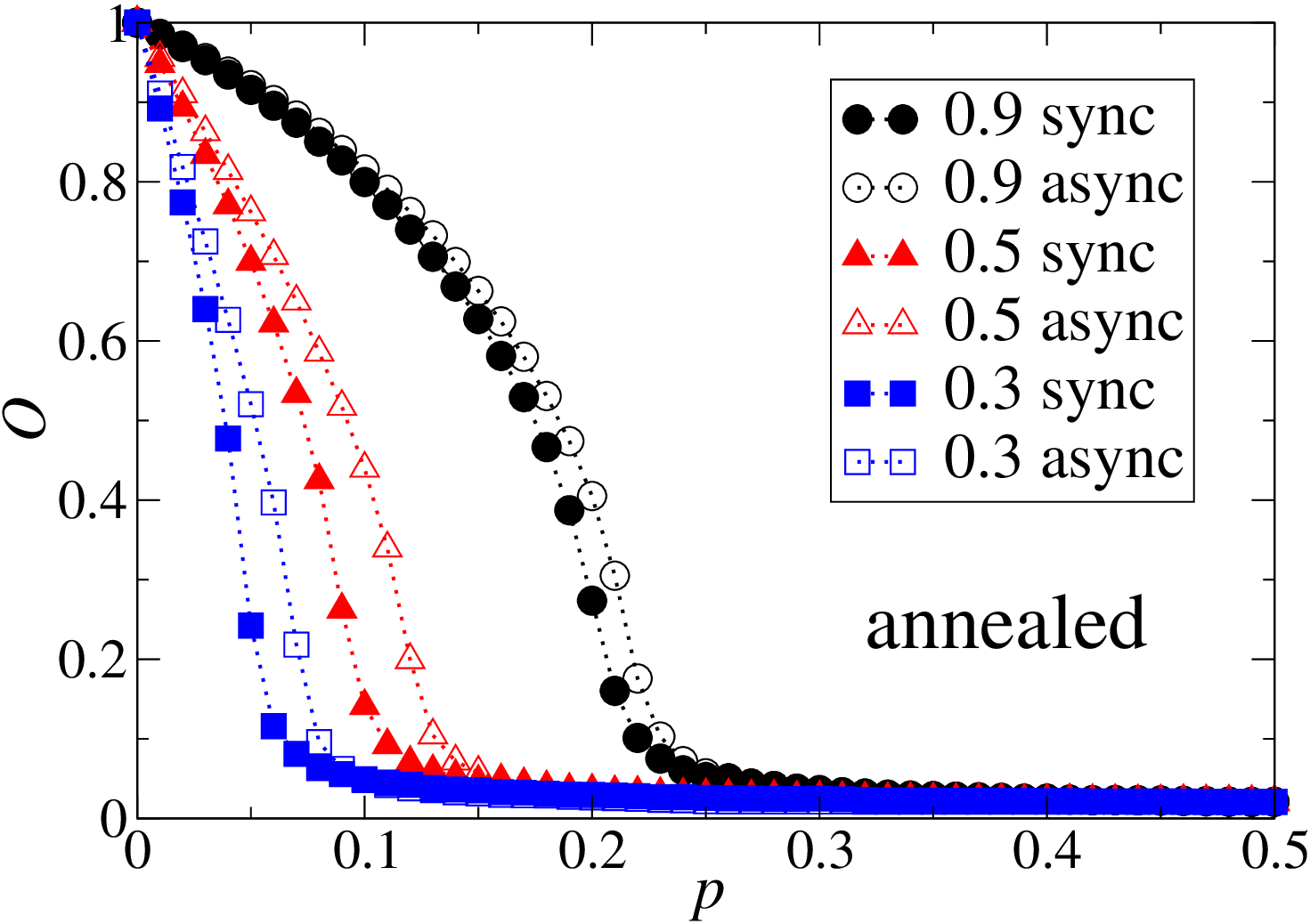}
\end{center}
\caption{(Color online) 
Order parameter versus $p$ for quenched (top) and annealed (bottom) 
random variables  of Eq. (\ref{Cdist1}), with typical values of $q$ indicated on the figure. 
The full (open) symbols are results of simulations with synchronous (asynchronous) update. 
The population size is $N=1000$ and data are averaged over $100$ realizations.}
\label{fig:Op1}
\end{figure}

For the distribution $G_1(C_i)$ of Eq. (\ref{Cdist1}), the mean fraction of null  convictions 
is  $1-q$. Such agents with no convictions  evolve influenced only by the 
interaction with other randomly chosen agents.


\begin{figure}[b]
\begin{center}
\vspace{0.4cm}
\includegraphics[width=0.4\textwidth,angle=0]{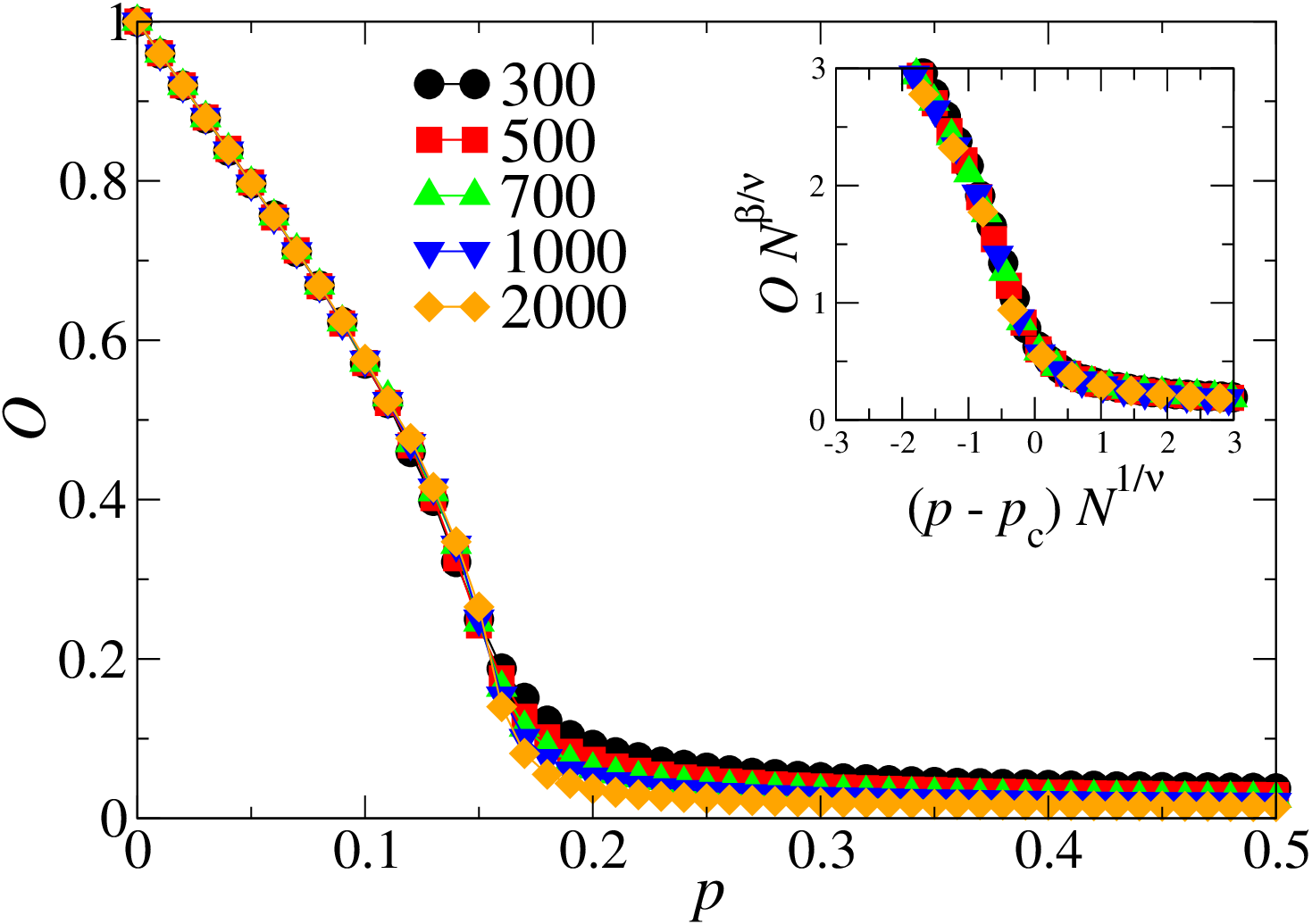}\\
\vspace{0.6cm}
\includegraphics[width=0.4\textwidth,angle=0]{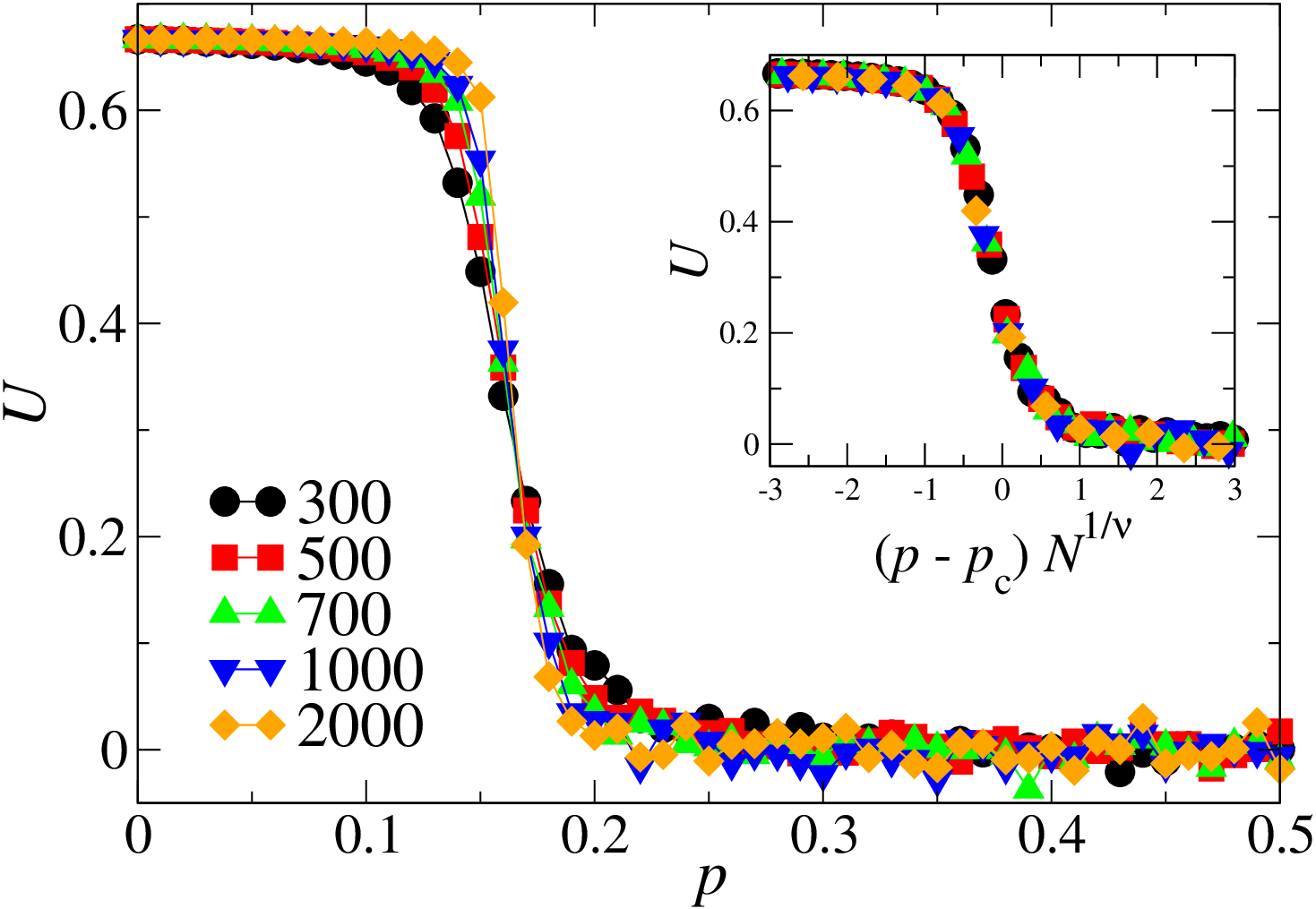} \\
\vspace{0.5cm}
\includegraphics[width=0.4\textwidth,angle=0]{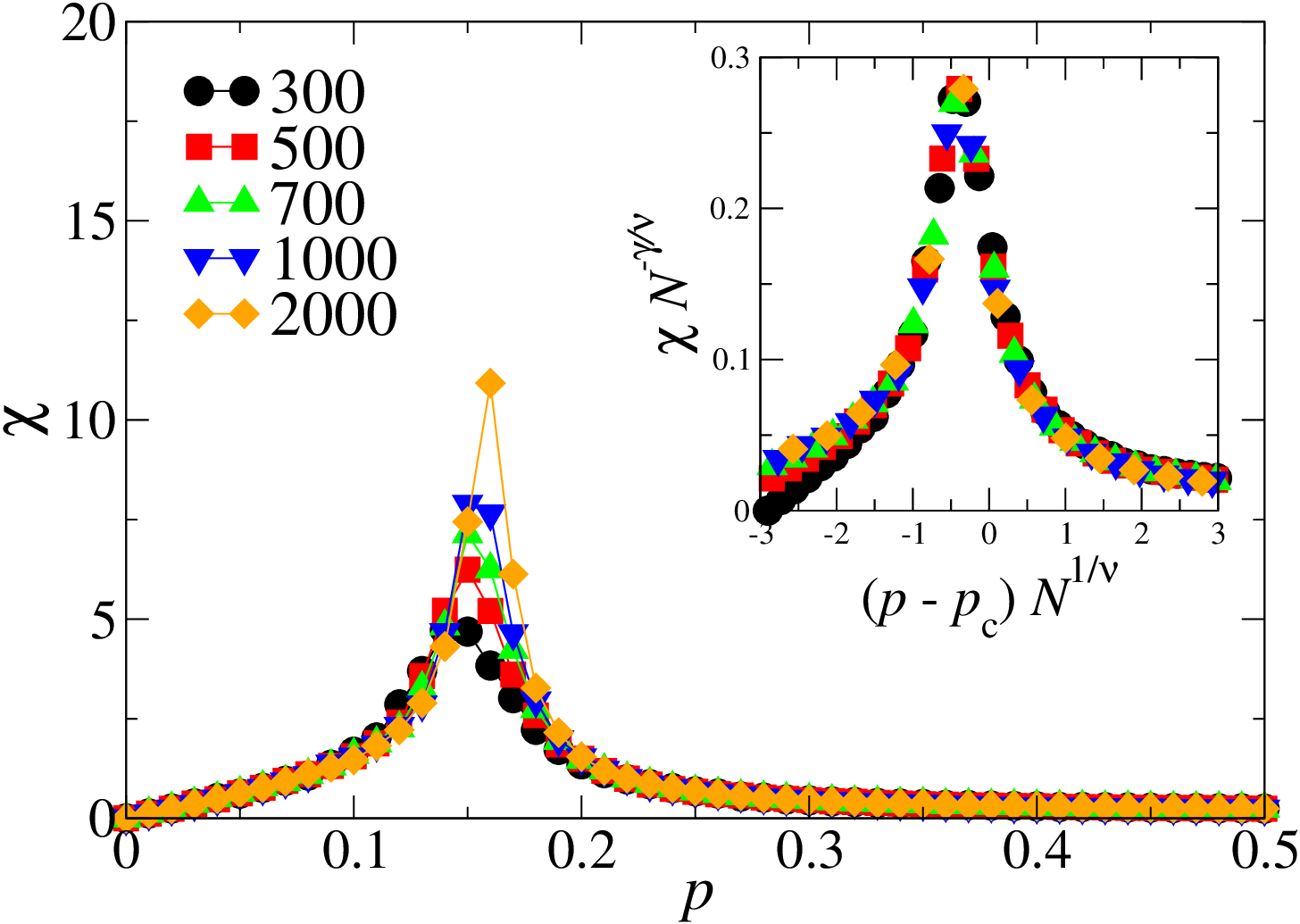}
\end{center}
\caption{(Color online) Order parameter, Binder cumulant and susceptibility   
for the PDF of Eq. (\ref{Cdist1}) with $q=0.5$ and different population sizes $N$ indicated 
on the figure (main plots). 
The corresponding scaling plots are shown in the respective insets. 
Data are for quenched random variables and synchronous update scheme. 
The best data collapse was obtained for $p_c\sim 0.167$, 
$\beta\sim 0.5$, $\gamma\sim 1.0$ and $1/\nu\sim 0.5$.}
\label{fig:scaling1}
\end{figure}

In Fig. \ref{fig:Op1} we exhibit results for the order parameter as a function of $p$ for typical values 
of $q$, allowing to compare the cases with quenched (top) and annealed (bottom) disorder. 
One can see that 
the curves for synchronous and asynchronous updates in the quenched case are almost identical, 
indicating that the critical behavior is not modified by the update scheme 
when we consider frozen disorder. On the other hand, 
if we allow the disorder to fluctuate in time, the results for synchronous 
and asynchronous updates are different.

We have verified numerically that, in all the analyzed cases of disorder and 
update scheme, the system undergoes a non-equilibrium 
phase transition for all values of $q>0$. 
The transition separates an ordered phase, where one of the extreme opinions 
($+1$ or $-1$) dominates, from a disordered one, where the three opinions coexist 
equally. A condition that was also obtained analytically  for the 
synchronous annealed case (see the Appendix).  
 
 In order to locate the critical points $p_{c}(q)$ numerically, we have performed simulations 
for different population sizes $N$. Thus, the transition points $p_{c}(q)$ are estimated, 
for each value of $q$, from the crossing of the Binder cumulant curves for the different sizes. 
In addition, a finite-size scaling analysis was performed, in order to obtain an estimate 
of the critical exponents $\beta$, $\gamma$ and $\nu$. As an illustration, we exhibit in 
Fig. \ref{fig:scaling1} the behavior of the quantities of interest as well as the scaling plots 
for $q=0.5$, quenched random couplings and synchronous updates. 
Our estimates for the critical exponents coincide with 
those for the original model ($q=1$), i.e., we obtained $\beta\sim 0.5$, $\gamma\sim 1.0$ 
and $1/\nu\sim 0.5$. These exponents are robust: they are the same for all values 
of $q$, independently of the update scheme considered and of the kind of random variables 
(quenched or annealed).

\begin{figure}[h!]
\begin{center}
\vspace{0.6cm}
\includegraphics[width=0.4\textwidth,angle=0]{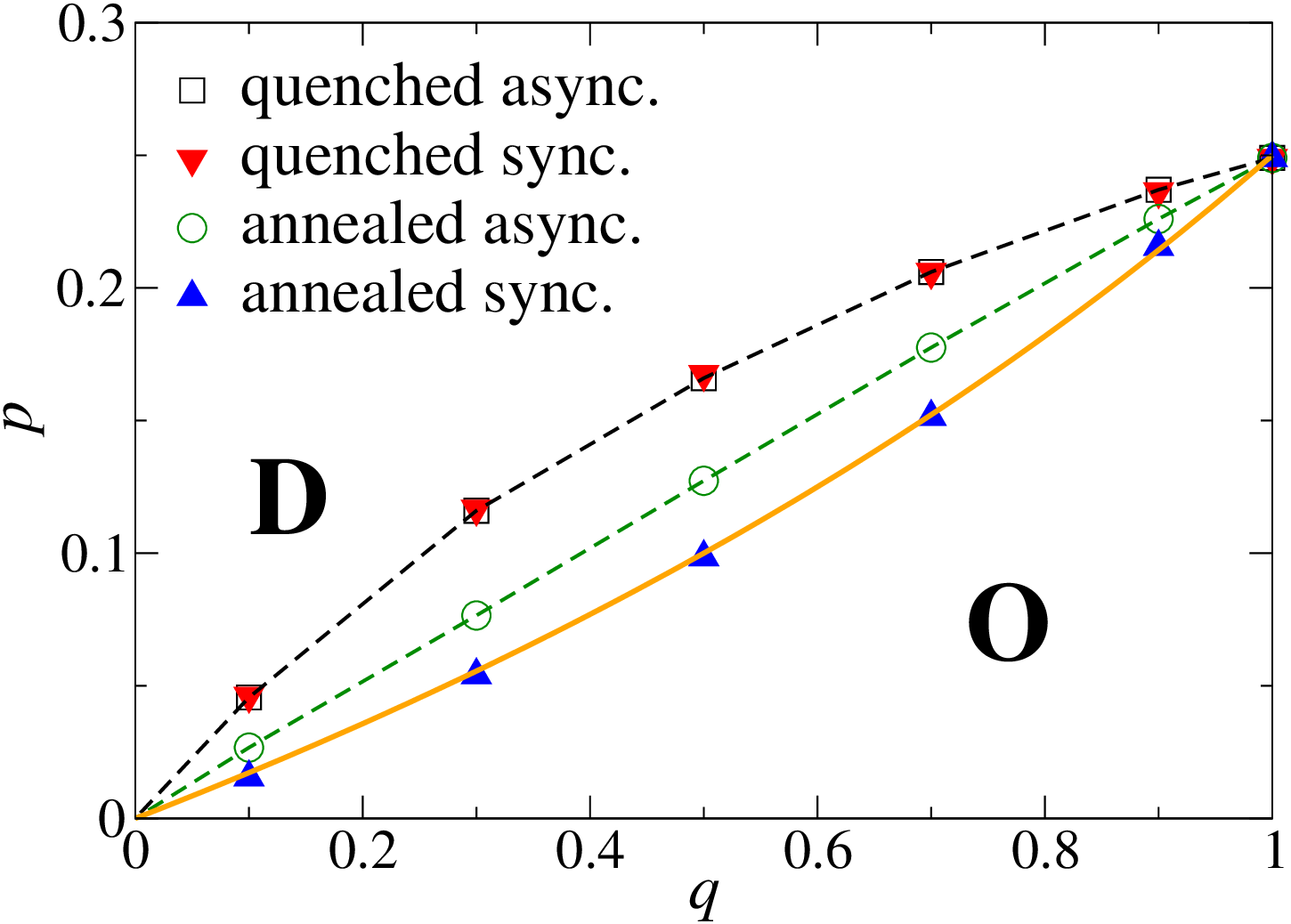}
\end{center}
\caption{(Color online) Phase diagram of the model defined by 
Eqs. (\ref{ev_eq}), (\ref{mudist}) and (\ref{Cdist1}) 
in the plane $p$ versus $q$, separating the ordered (\textbf{O}) and disordered (\textbf{D}) 
phases. The symbols are the finite-size estimates of the critical points $p_{c}(q)$ obtained from 
the simulations, dashed lines are  guides to the eye  and the full line 
is the analytical result  predicted by Eq. (\ref{an_syn01}).
}
\label{fig:diagram1}
\end{figure}

Taking into account the critical values $p_{c}(q)$ obtained from the simulations, 
we exhibit in Fig. \ref{fig:diagram1} the phase diagram of the model in the plane $p$ versus $q$. 
As discussed before, in the case of quenched variables the frontier is independent of 
the update scheme. On the other hand, for annealed variables the results are 
different. This is possibly due to the time fluctuation of the annealed variables, 
which does not occur  in the quenched case. 
The analytical prediction for the synchronous annealed case is presented in the Appendix.

\subsection{Model with distribution $G_2$}

For the distribution $G_2(C_i)$ of Eq. (\ref{Cdist2}), a fraction $1-q$ of the convictions 
$C_i$ are $-1$ (instead of being null as in Sec. III.A). 
Now, some agents $i$ present negative convictions that contribute to a spontaneous change in their opinions, together with the influence of a randomly chosen agent $j$.

\begin{figure}[b!]
\begin{center}
\vspace{0.7cm}
\includegraphics[width=0.45\textwidth,angle=0]{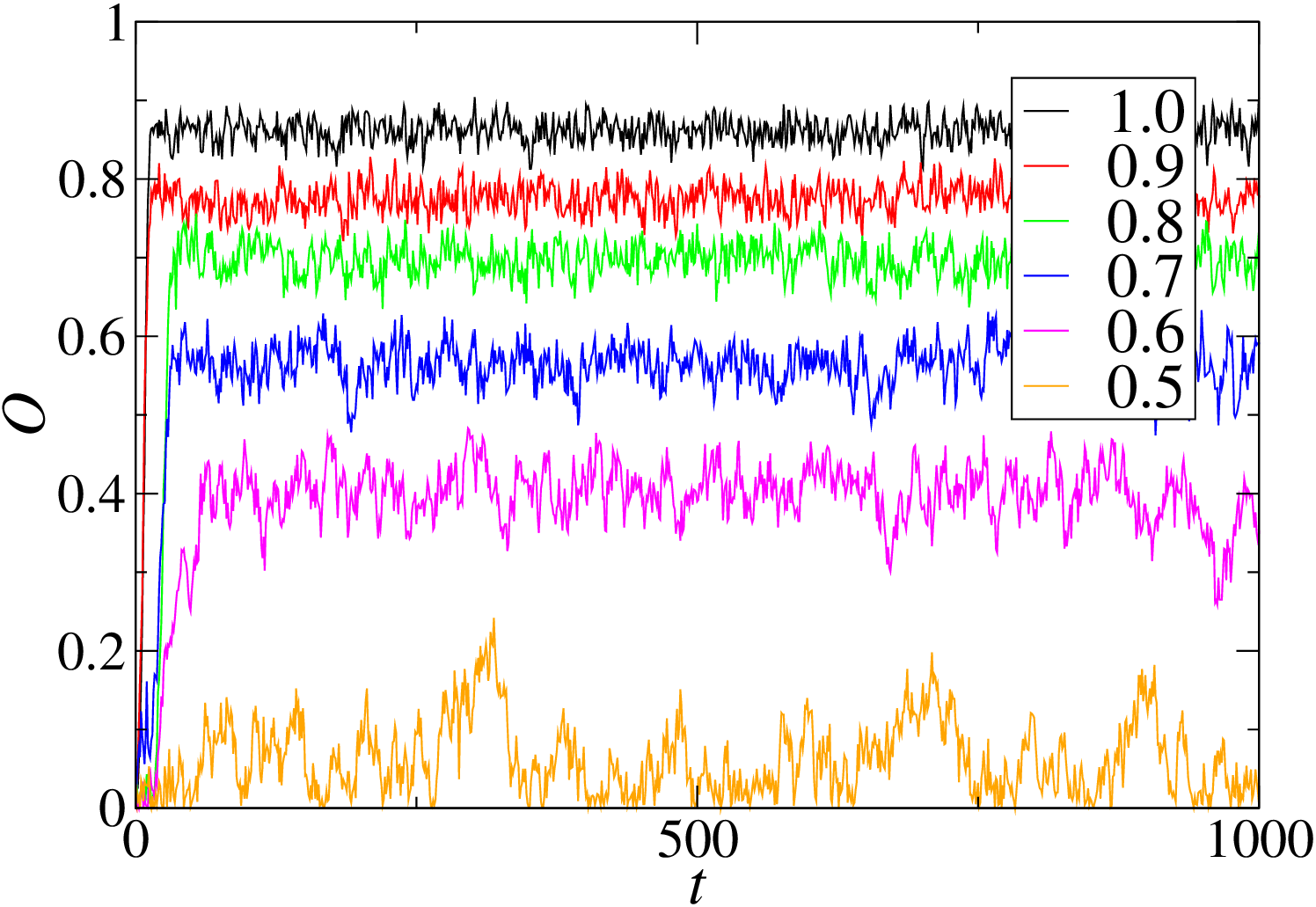} \\
\vspace{0.57cm}
\includegraphics[width=0.45\textwidth,angle=0]{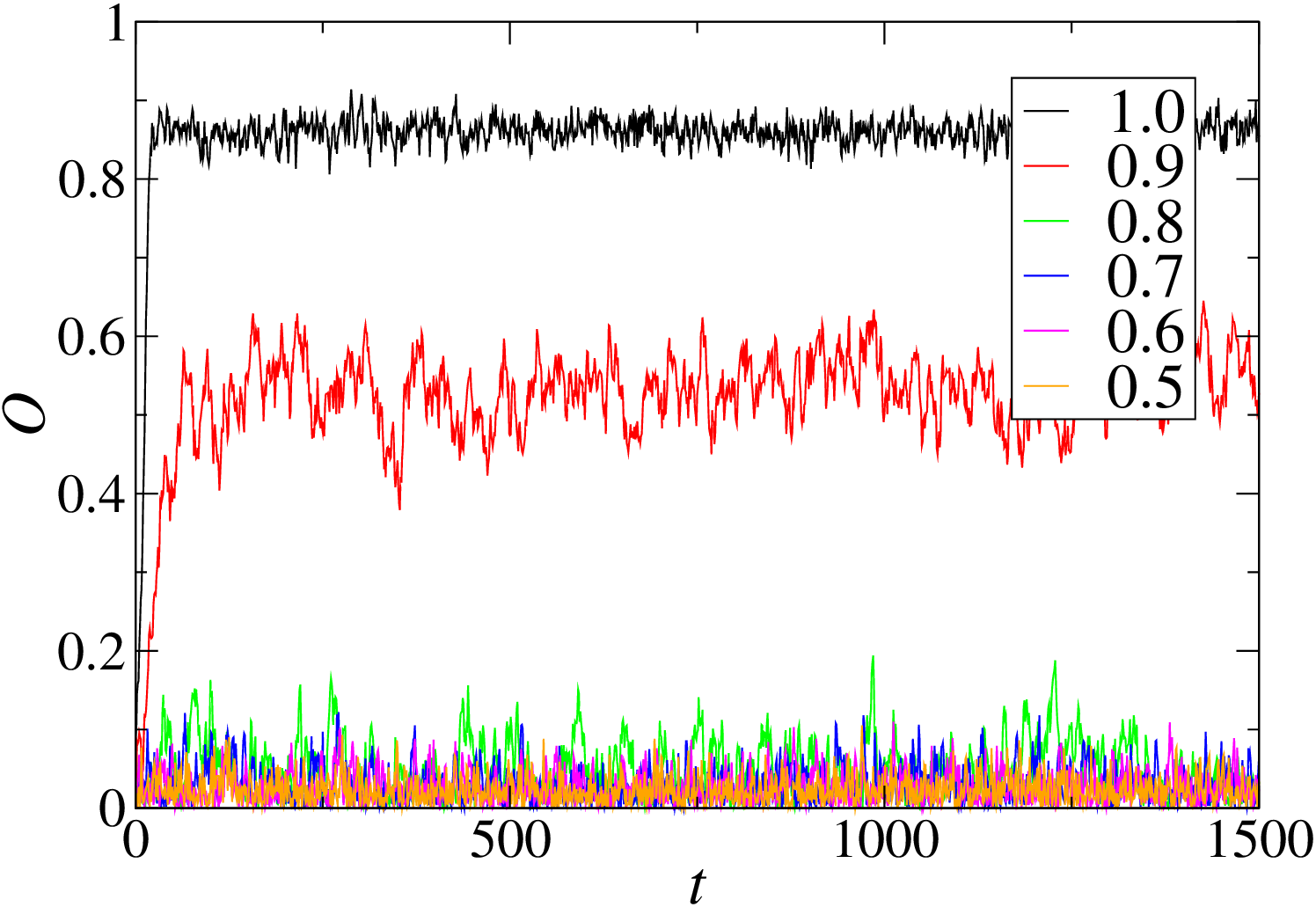}
\end{center}
\caption{(Color online) Time evolution of the order parameter for $N=1000$, $p=0.1$ and 
typical values of $q$, labeling the curves from top to bottom. 
The curves are for:   quenched asynchronous (top)  and  annealed synchronous (bottom). 
%
We can observe that a disordered state is reached when the value of $q$ is decreased 
below a threshold.}
\label{fig:evolution}
\end{figure}

Differently from the case where the PDF of convictions is given by Eq. (\ref{Cdist1}), 
analyzed in Sec. III.A, now we can observe that there is a threshold $q_{c}$ below which the 
system is always in a disordered state, for all values of $p$. 

The time evolution of the order parameter is illustrated in Fig.~\ref{fig:evolution} 
for the quenched asynchronous and annealed synchronous cases. Similar evolution is observed 
for the other two combinations too. For sufficiently low $q$, none of the two 
extreme opinions dominates. Moreover, we verified that in such cases 
the fraction of each one of the three possible opinions is again $1/3$ in average, indicating complete disorder. 
This result was also found theoretically for the annealed synchronous case (see Appendix).

In the cases where a transition occurs,  a finite-size scaling analysis 
was perfomed as in Sec. III.A.  
The same mean-field exponents were obtained, independently of the update 
scheme considered and of the kind of random variables.
The phase diagram for the different types of update and random variables is shown in 
Fig.  \ref{fig:diagram2}. 
The analytical prediction for the synchronous annealed case, derived  in the Appendix, 
is also included. 
In such case, Eq. (\ref{an_syn11}) predicts that for $q<3/4$ no transition occurs, and 
the system is always in a disordered state.

\begin{figure}[h]
\begin{center}
\vspace{0.6cm}
\includegraphics[width=0.4\textwidth,angle=0]{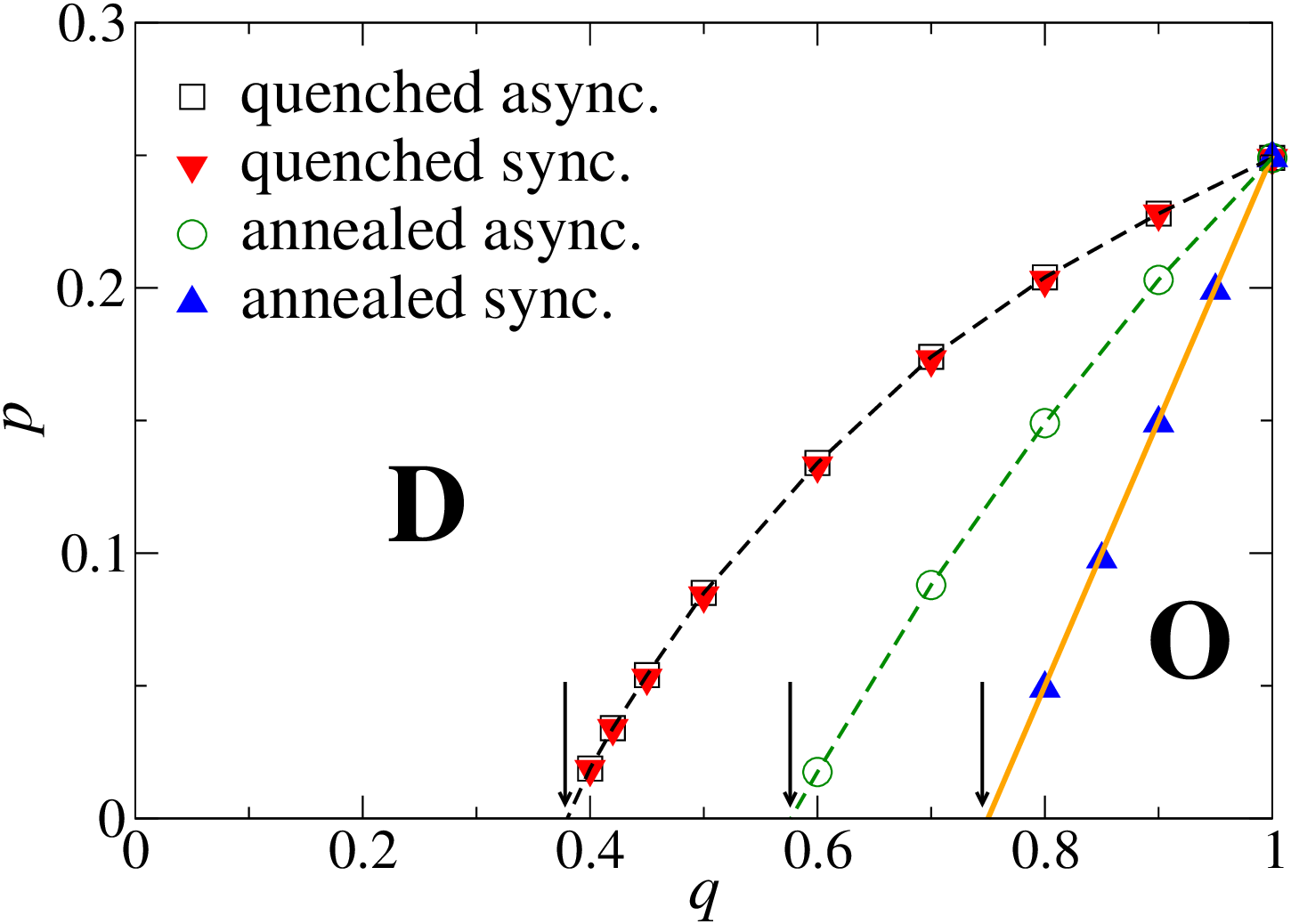}
\end{center}
\caption{(Color online) Phase diagram of the model defined by Eqs. (\ref{ev_eq}),  
(\ref{mudist}) and (\ref{Cdist2}) in the plane $p$ versus $q$, separating the ordered (\textbf{O}) 
and disordered (\textbf{D}) phases. The symbols are the finite-size estimates of the critical points 
$p_{c}(q)$ obtained from the simulations, the dashed lines are guides to the eye, 
and the full line is the analytical result given by Eq. (\ref{an_syn11}). 
Notice that the transition is destroyed for values of $q$ below a threshold, indicated schematically by arrows. 
}
\label{fig:diagram2}
\end{figure}


\section{Final Remarks}

In this work we have studied opinion dynamics through a model where agents interact 
by pairs in a fully-connected graph. 
The opinions have three different states (spin-1) and 
the agents interact through random couplings that can be 
positive/ferromagnetic (or negative/anti-ferromagnetic) 
with probability $1-p$ (or $p$). 
Differently from other related models, 
where the ordered state is marked by consensus, the ordered state 
is characterized by the upraise of a dominating extreme opinion, which becomes consensual 
only in the limit in which interactions are all positive ($p=0$).
Moreover, there is also a self-interaction term, the conviction, 
which we considered to assume random values, according to the PDFs $G_1$ or $G_2$, 
given by Eqs. (\ref{Cdist1}) and (\ref{Cdist2}), respectively. 
Then, we aimed to study the impact of the heterogeneity of convictions 
on the critical behavior of opinion formation. 
Although states and couplings can take only a few values, 
a wider spectrum of possibilities is expected to be somehow mapped on the present simpler case.

First, we have considered the PDF $G_1$ that aims to model populations where there 
is a fraction $1-q$ of agents without self-convictions about their opinions, 
and thus they can be easily persuaded to change their opinions. 
Our results show that the critical fraction of negative interactions 
$p_{c}$ below which the population reaches partial agreement decreases smoothly 
for decreasing values of $q$, collapsing with $p_{c}=0$ only at $q=0$. 
This order-disorder transition is continuous and the critical exponents 
are universal and mean-field like, presenting the values $\beta=0.5$, 
$\gamma=1.0$ and $1/\nu=0.5$ for all values of $q$.

We have also considered the PDF $G_2$ for the convictions in societies 
where some agents have a tendency to change spontaneously their opinions. 
In this case, disordered states 
are favored, and the order-disorder boundary 
falls off rapidly to $p_{c}=0$ for decreasing values of $q$. 
Thus, in opposition to the previous case, there are threshold values of $q$ 
below which the system is always in the disordered state. 
Despite this difference, the order-disorder transition is also continuous 
and the critical exponents are universal and mean-field like, 
as in the previous case. 
 
Notice that the introduction of negative interactions, pondered by the probability $p$,  
produces a similar effect of the so-called Galam's contrarians \cite{galam_cont,delalama}. 
In fact, in the absence of negative couplings ($p=0$)
the system presents consensus states with one of the extreme opinions ($+1$
or $-1$) dominating the population. On the other hand, the inclusion of a
fraction of negative interactions leads the system to a disordered state
with the coexistence of the three possible opinions $+1$, $-1$ and $0$,
analogous to the stalemate state produced by the introduction of
contrarians in opinion models, where the two possible opinions, namely $+1$
and $-1$, coexist \cite{galam_cont,delalama}. Observe also that the introduction of the
conviction parameter $q$ makes this effect more pronounced. In fact, the
critical values $p_{c}$ decrease for increasing values of $q$, and in the
case of the bimodal distribution $G_2$, the effect of the convictions is so
strong that it destroys the order-disorder transition.

It is important to notice that the results depend quantitatively (but no qualitatively) 
on the kind of update scheme used (synchronous or asynchronous) and on the nature 
(quenched or annealed) of the random variables considered, for the two studied PDFs.

\appendix
\section{}
\label{app}

Following the lines of Ref.~\cite{biswas}, we computed critical values for the 
synchronous annealed case. 
We first obtained the matrix of transition probabilities  
whose elements $m_{i,j}$ furnish the probability that a state suffers the 
shift or change $i \to j$.
Let us also define $f_1$, $f_0$ and $f_{-1}$, the stationary probabilities of each possible state. 

In the steady state, the fluxes into and out from a given state must balance. 
In particular, for the  null state, one 
has $m_{1, 0}+m_{-1, 0}=m_{0,1}+m_{0,-1}$. 
Moreover, when the order parameter vanishes, it must be $f_1=f_{-1}$. 
In both cases considered below for the distribution of convictions, 
those two equalities imply $f_1=f_{-1}=f_0=1/3$ (disorder condition). 
This holds in particular at the critical point.

Finally, let us define $r(k)$, with $-2\le k \le 2$, the 
probability that the state shift per unit time is $k$, that is, $r(k)=\sum_i m_{i,i+k}$.  
In the steady, the average shift must vanish, namely, 
\begin{equation} \label{nullshift}
 2[r(2)-r(-2)] +r(1)-r(-1)=0 \,.
\end{equation}

\subsection{PDF $G_1$}
\label{ap1}

 The elements of the transition matrix are
 
\begin{eqnarray} \nonumber
m_{1, 1} &=& f_1^2 (1 - p) + f_1 f_0 q + f_1 f_{-1} p \\ \nonumber
m_{1, 0} &=& f_1^2 q p + f_1 f_0 (1 - q) + f_1 f_{-1} q (1 - p)\\ \nonumber
m_{1, -1} &=& f_1^2 (1 - q) p  + f_{-1} f_1 (1 - q) (1 - p)\\ \nonumber
m_{0, 1} &=& f_0 f_1 (1 - p) + f_0 f_{-1} p\\ \nonumber
m_{0, 0} &=& f_0^2\\ \nonumber
m_{0, -1} &=& f_0 f_1 p  + f_0 f_{-1} (1 - p)\\ \nonumber
m_{-1, 1} &=& f_1 f_{-1} (1 - q) (1 - p) + f_{-1}^2  p (1 - q) \\ \nonumber
m_{-1, 0} &=&  f_1 f_{-1} q (1 - p) + f_0 f_{-1}  (1 - q) + f_{-1}^2 q p\\ \nonumber
m_{-1, -1} &=& f_1 f_{-1} p + f_0 f_{-1} q  + f_{-1}^2 (1 - p)\,. 
\end{eqnarray}
 
The null average shift condition (\ref{nullshift}), together with the disorder condition, 
leads to 
 
\begin{equation} \label{an_syn01}
p_c = \frac{q}{2(3-q)} \,.
\end{equation}

\subsection{PDF $G_2$}
\label{ap2}

For this PDF, the transition matrix is  \\[4mm]
  
\begin{eqnarray} \nonumber
m_{1, 1} &=& f_1^2 q(1 - p) + f_1 f_0  q + f_1 f_{-1}  q p  \\ \nonumber
m_{1, 0} &=& f_1^2 (q p + (1 - q)(1 - p)) + f_1 f_{-1} (p + q - 2 p q)\\ \nonumber
m_{1, -1} &=& f_1^2 (1 - q)p  + +f_1 f_0 (1 - q) + f_{-1} f_1 (1 - q)(1 - p)\\ \nonumber
%
m_{0, 1} &=& f_0 f_1 (1 - p) + f_0 f_{-1} p \\ \nonumber
m_{0, 0} &=& f_0^2 \\ \nonumber
m_{0, -1} &=& f_0 f_1 p  + f_0 f_{-1}(1 - p) \\ \nonumber
m_{-1, 1} &=& f_1 f_{-1} (1 - q)(1 - p) + f_{-1} f_0 (1 - q) + f_{-1}^2  p(1 - q) \\ \nonumber
m_{-1, 0} &=&  f_1 f_{-1}  (p + q - 2 p q)  + f_{-1}^2( (1 - p)(1 - q) + q p) \\ \nonumber
m_{-1, -1} &=& f_1 f_{-1} p  q + f_0 f_{-1} q  + f_{-1}^2 q(1 - p) \,.
\end{eqnarray}

In this case Eq. (\ref{nullshift}), toghether with the disorder condition, gives
\begin{equation}  \label{an_syn11}
p_c=q-3/4 \,.
\end{equation}
 
In contrast to  the frontier defined by Eq. (\ref{an_syn01}), which implies a critical value of 
$p$ below which the system has a predominant opinion, 
Eq. (\ref{an_syn11}) implies that for $q<3/4$ the system can not order.

\section*{Acknowledgements:}
The authors are grateful to Thadeu Penna for having provided the computational resources of the Group of Complex Systems of the Universidade Federal Fluminense, Brazil, where the simulations were performed. This work was supported by the Brazilian funding agencies FAPERJ, CAPES and CNPq.

\end{document}